\documentclass[preprint,secnumarabic,amssymb, nobibnotes, aps, prd]{revtex4-1}

\usepackage[pdftex]{graphicx}  
\begin{document}

\title{Modifications of filament spectra by shaped octave-spanning laser pulses}%
\author{A. Patas $^1$}
\author{M. Matthews $^2$}
\author{S. Hermelin $^{2,3}$}
\author{J. Gateau $^2$}
\author{J. Kasparian $^{2,4}$}
\author{J. P. Wolf $^2$}
\author{A. Lindinger $^1$}
\email[Corresponding author: ]{lindin@physik.fu-berlin.de}
\affiliation{$^1$ Institut f\"{u}r Experimentalphysik, Freie Universit\"{a}t Berlin, Arnimallee 14, 14195 Berlin, Germany}
\affiliation{$^2$ Group of Applied Physics, Universit\'{e} de Gen\`{e}ve, Chemin de Pinchat 22, 1211 Geneva 4, Switzerland}
\affiliation{$^3$ University of Lyon, Universit\'{e} Claude Bernard Lyon 1, CNRS, Institute Lumi\`{e}re Mati\`{e}r, 690622 Villeurbanne, France}
\affiliation{$^4$ Institute for Environmental Sciences, Universit\'{e} de Gen\`{e}ve, Boulevard Carl Vogt 66, 1205 Geneva 4, Switzerland}

\date{June 7, 2018}%

\begin{abstract} 
In this paper we examine the spectral changes in a white light laser filament due to different pulse shapes generated by a pulse shaping setup. We particularly explore how the properties of the filament spectra can be controlled by parametrically tailored white light pulses. The experiments are carried out in a gas cell with up to $9\,bar$ of argon. Plasma generation and self-phase modulation  strongly affect the pulse in the spectral and temporal domain. By exploiting these effects we show that the pulse spectrum can be modified in a desired way by either using second order parametric chirp functions to shift the filament spectrum to higher or lower wavelengths, or by optimizing pulse shapes with a genetic algorithm to generate more complex filament spectra. This paper is one of the first examples of the application of complex, parametrically shaped white light pulses.
\end{abstract}

\maketitle

\section{Introduction}

A broad spectrum is required when generating ultrashort laser pulses for 
exploring very fast molecular processes. Flamentation in gas \cite{chin2005,berge2007,cou2007} or in hollow-core 
fibers \cite{wadsworth_supercontinuum_2002, zheltikov_let_2006} is particularly utilized for spectral broadening to generate few cycle white light laser pulses
with high time resolution. Such short pulses were for example applied to obtain electronic
wave packets \cite{li_coherent_2015}. Tailoring of laser pulses is another current achievement with high potential since by using longer shaped pulses, it was already demonstrated that excitation
pathways can be selected so that ionization \cite{wollenhaupt2002interferences, lindinger2004isotope},
dissociation \cite{assion1998control, kling_control_2006}, or 
fluorescence  \cite{roth2009quantum} can be controlled. The optimal pulse shapes were often found in
a closed loop scheme.
Recently, pulse shaping was  for the first time
used with white light laser pulses from filaments in gases to optimize the ionization ratio of small silver clusters
\cite{hagemann_2015} and for the energy deposition in rare gases \cite{gat2018}.

Laser filamentation in gases is a fascinating process involving dispersion, multiple nonlinear and spatial
effects, as well as ionization dynamics. For
the necessary cancellation of self-focusing and plasma
defocusing high peak intensities are required, either by using high powered lasers or extremely short pulses. For many applications a white light supercontinuum
is generated by using sapphire plates
\cite{imran2010efficient, jarman2005supercontinuum}. In contrast, we utilize an intense Ti:Sa laser with
a two-stage flamentation in air to create the broad
spectrum \cite{matt2018,stein1999}. This approach has the advantage of providing
comparatively high pulse energies. 
Ionization processes depend on the maximal pulse intensity, the wavelength and the ionization potential.
The Keldysh parameter  \cite{Keldysh1965} is relevant to decide whether ionization follows mainly
multi-photon ionization (MPI) or tunnel-ionization (TI).
Theories were developed  (ADK \cite{ammosov_tunnel_1987}, PPT \cite{perelomov_ionization_1966}) in order to describe the ionization rates \cite{talebpour1999semi}. They enabled the simulation of filamentation
and laser pulse propagation and provided insight into
the dynamics resulting from
spatial and temporal effects. Spectral blue shifting \cite{le_blanc_spectral_1993},
temporal pulse breaking \cite{bree2010plasma} and filamentation over long
distances  \cite{durand_kilometer_2013} have been observed. Laser filamentation is
highly nonlinear and sensitively relies on
the incoming pulse shape. Recent investigations of the influence 
of linear chirp and of the temporal and
spatial focusing enabled the shift and control of
the spatial onset of the filament  \cite{kasparian_white-light_2003} and the pulse shaping of the filaments \cite{heck2006,acker2006,zah2014}. 

In this paper
we want to study the influence of the shape of ultrabroadband pulses on
the resulting spectra after filamentation in argon.
Our unique laser system with pulse shaping unit allows
us to arbitrarily change the phase and amplitude of
octave-spanning white light laser pulses  \cite{Hagemann2013}. 
The spectra at the filament output will provide insight into
the interplay of the occuring effects and provide guidance to get various spectral forms by tuning the initial pulse parameters. This contribution is partially received from the doctoral thesis of A. Patas \cite{Patas}. 

\section{Experimental Setup}

\begin{figure}[!]
	\centering
  \includegraphics[width=0.5\textwidth]{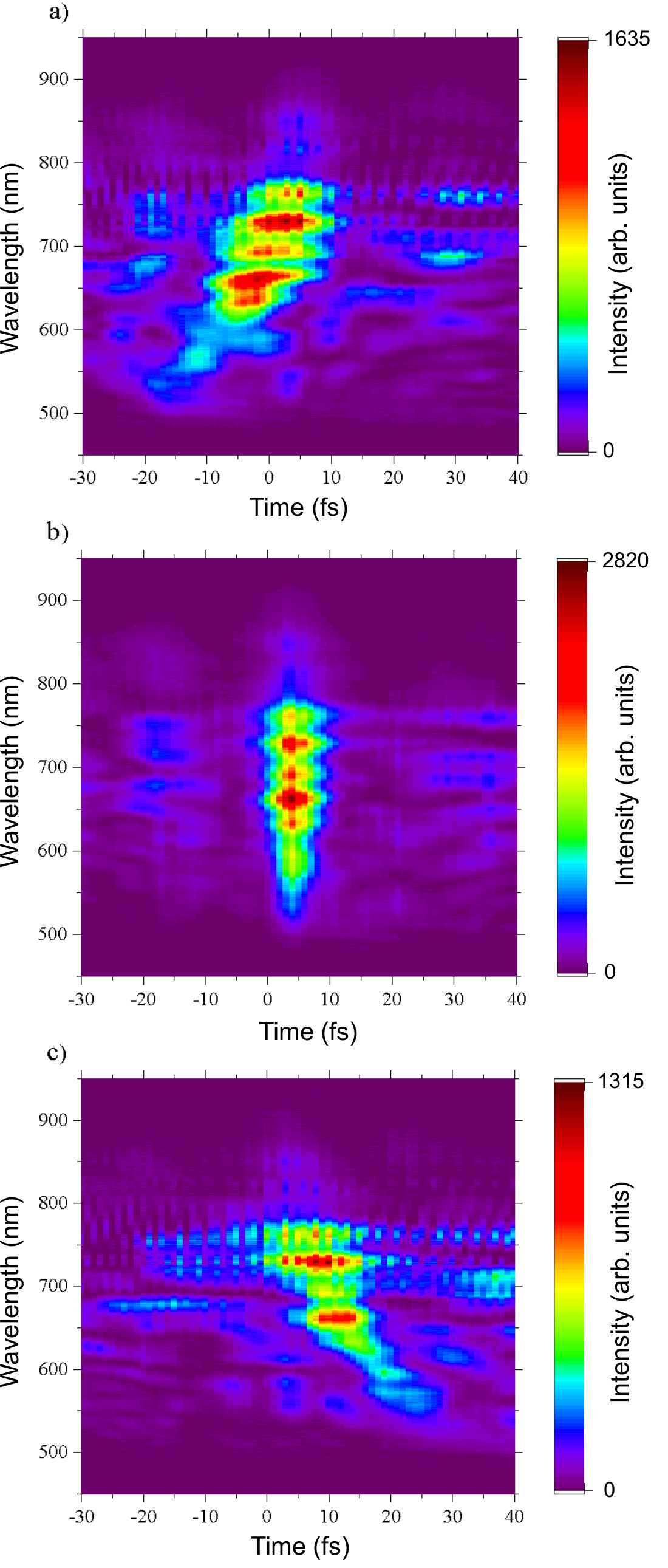}
	\caption{TG-FROG traces of chirped and unchirped laser pulses before the argon chamber. a) Negatively chirped pulse with $-20\,fs^2$. b) Transform-limited pulse. c) Laser pulse which is positively chirped with $20\,fs^2$.  Higher wavelengths are underrespresented due to the calibration of the TG-FROG.}
	\label{TG-Frog}
\end{figure}

The laser system used in this experiment is capable of producing few cycle, white light laser pulses and was extensively described earlier \cite{Hagemann2013}. It consists of a Ti:Sa oscillator (Femtosource Compact, Spectra-Physics) followed by a multipass amplifier (Odin C, Quantronix) resulting in pulses at a rate of $1\,kHz$ with a central wavelength of $800\,nm$ and a spectral width of about $100\,nm$. The laser beam diameter amounts to 8$\pm0.3$ $mm$. The laser power is reduced using an aperture to $490\,mW$ which has shown to give the optimal spectral broadening after a first $f=4\,m$ filamentation stage in air. The spectrally broadenend and chirped pulses are then compressed by a pair of broadband chirped mirrors. Then a second $(f=3\,m)$ filamentation is used to further broaden the laser spectrum to the octave-spanning result. The pulses are again compressed with another pair of chirped mirrors before entering the pulse shaping setup. The folded 4f setup uses cylindrical focusing mirrors instead of lenses to avoid chromatic abberation. We use a pulse shaper (SLM640, CRI) with two liquid crystal arrays of 640 pixels each. These arrays have their optical axes oriented at plus and minus 45 degrees with respect to the laser polarization. This allows for phase and amplitude modulation of the complete white light spectrum. Extensive tests and calibration were required to ensure a correct phase and amplitude response at each wavelength. A wire-grid polarizer is used after the pulse shaping setup to allow for amplitude modulation if required.

\begin{figure}[ht]
	\centering
  \includegraphics[width=0.8\textwidth]{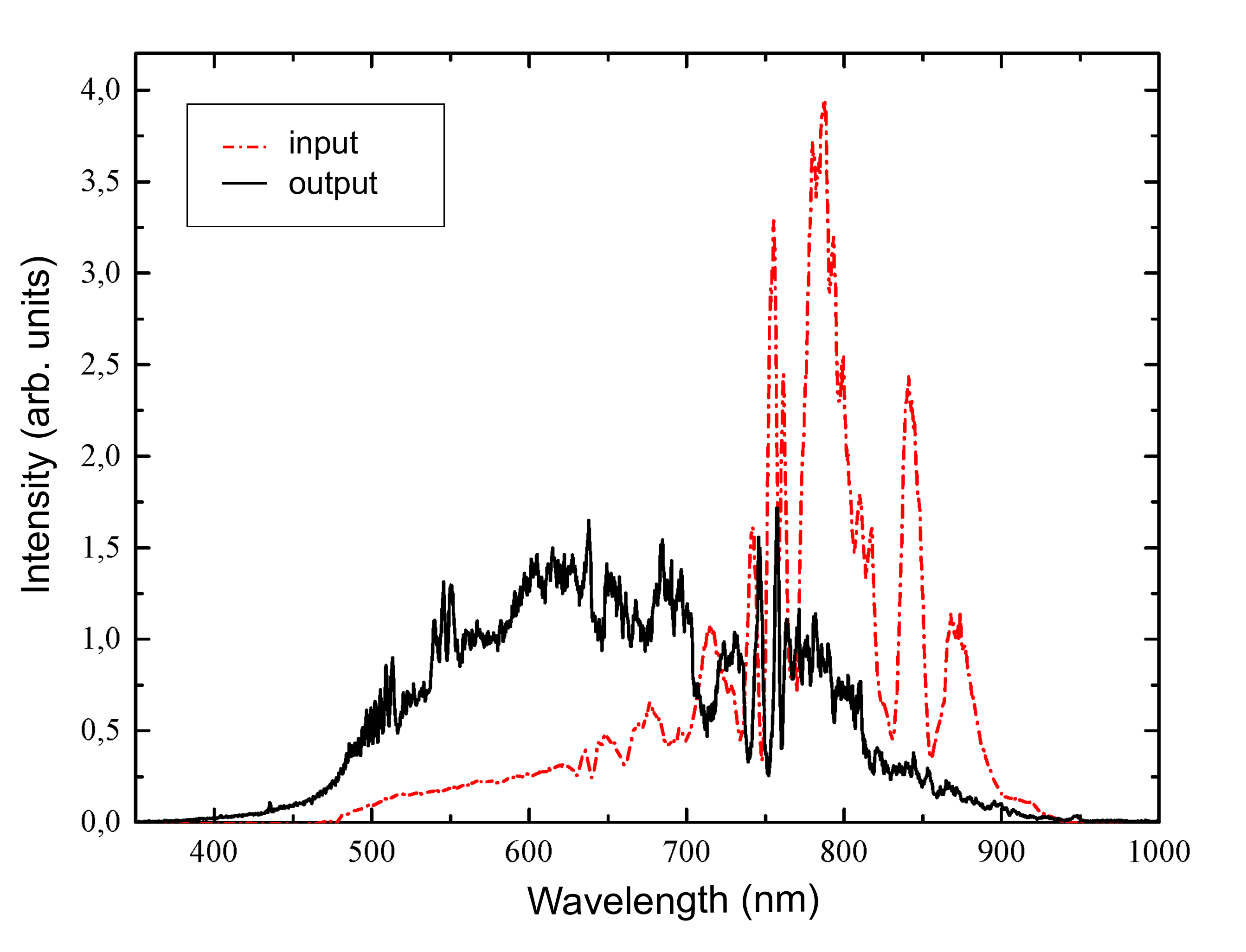}
	\caption{Laser reference spectrum used for the experiments (dashed red line). The spectrum was recorded directly at the entrance of the chamber filled with 9 bar of argon. The strongly modified spectrum for the same  transform-limited laser pulse   after transmission through the argon chamber is also shown (solid black line).}
	\label{TLvsRef}
\end{figure}

We were able to verify the performance of the setup with the help of a TG-FROG \cite{Hagemann2013}. Using the pulse shaper to compensate for the remaining pulse-chirp we repeatedly measured TG-FROG lengths of $5-7\,fs$ (FWHM) with pulse energies up to $50\,\mu J$ just before the measurement chamber. Since the TG-FROG is very sensitive to the peak-intensity, we were not able to measure pulses with more complex shapes. As a demonstration of the capability of the setup Fig. \ref{TG-Frog} shows positive and negative linearly chirped pulses as well as the short pulse as a reference. The corresponding laser spectrum is shown in Fig. \ref{TLvsRef} as a dashed red  line. After a couple of beam steering mirrors the beam is focused by a $250\,mm$ off axis focusing mirror within a gas chamber which is filled with 9 bar of argon (for details see \cite{matt2018}). The spectrum was measured after the laser beam passed the exit window of the chamber and hit a Teflon beam block with the spectrometer (USB2000 UV/VIS, Ocean Optics) pointing at the beam block. This method enables to detect the entire spectrum. Teflon is used because it reduces the amount of interference from multiple reflecting surfaces. 

The chirp induced by the gas as well as the entrance windows had to be accounted for each day, since small variations measurably changed the obtained pulse length. Therefore, an electret microphone was placed close to the filament position where it recorded the acoustic shock wave released by the pulse. This value is known to be correlated to the free carrier density produced by the laser pulse \cite{yu2003}. With the help of the PRISM \cite{Wu2011} algorithm the phase was optimized for the maximum acoustic shock before each experiment. Simulations show that this method is indeed a very suitable approach for finding the correct dispersion compensating offset phase at the onset of a filament.

\section{Results}

\begin{figure}[b]
	\centering
  \includegraphics[width=0.8\textwidth]{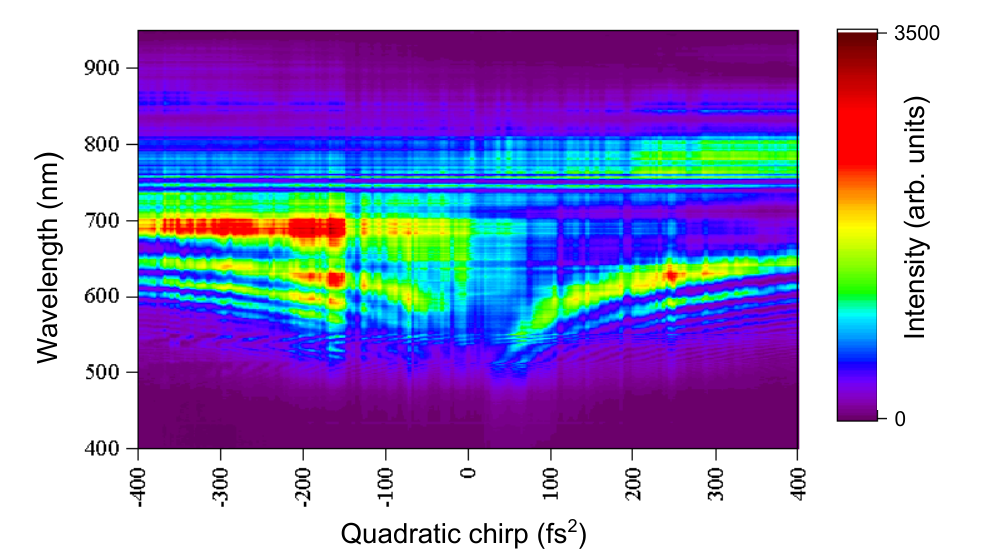}
	\caption{Spectra modified by nonlinear and plasma effects in the chamber. The x-axis denotes the quadratic chirp  of the pulses at the focal position. Strong spectral bands are observed which differ for positive and negative chirps. The asymmetry between positive and negative chirps is visible, resulting from the asymmetry of the pulse shapes.}
	\label{qscan}
\end{figure}

The dispersion compensating offset phase found by the PRISM algorithm was used to create precompensated pulses with the pulse shaper. These pulses are close to transform-limited at the onset of the filament. Fig. \ref{TLvsRef} (black line) shows the spectrum obtained after the chamber when using this offset phase. This phase compensates the dispersion of the gas and the optical elements until the position of the filament,
\begin{figure}[t]
	\centering
  \includegraphics[width=0.8\textwidth]{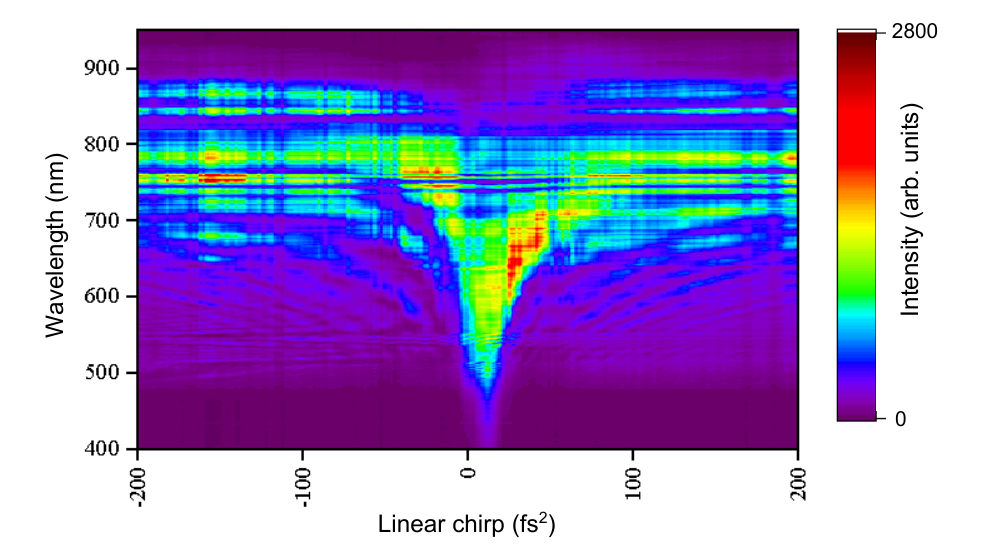}
	\caption{Contour plot of the spectra behind the chamber as vertical lines depending on the linear pulse chirp. A broadening of the spectra is obtained close to zero chirp and modulations occur in the low wavelength range for positive and negative chirps likewise.}
	\label{lscan}
\end{figure}
so that the spatial focus coincides with the temporal focussing of the laser pulse. Thereby the maximal peak intensity and ionization rate is reached in the beginning of the filament, resulting in the largest microphone signal. Dispersion phase compensation done at 3, 7 and 9 bar, with or without extra window in front of the chamber allowed us to find the exact dispersion matching one bar argon or a single chamber-window. The obtained  values are $610.5\,fs^2/m$ for GVD and $326.8 \, fs^3/m$ for GDD. The obvious pulse parameters to examine are the linear and quadratic chirp. Therefore spectra were taken while tuning the linear and quadratic chirp ($b_3$) from $-200\,fs^2$ to $200\,fs^2$ (Fig. \ref{lscan}) and $-400\,fs^3$ to $400\,fs^3$ (Fig. \ref{qscan}), respectively. The center wavelength for the chirp expansion was chosen to be $761.22\,nm$ which coincided with the center of the input spectrum.
It is known from former studies \cite{Sprangle2002,berti2015nonlinear} that only a combination of self-phase modulation, self-steepening and plasma effects leads to the measured spectra. The plasma acts as a defocussing element on the beam and thereby counteracts the self-focussing by the Kerr-effect. This was shown to result in repetitive focussing and defocussing \cite{mlejnek1998dynamic, talebpour1999re}.

The linear chirp scan shows modulations which shift to higher energies for smaller values in absolute chirp. 
Close to the transform-limit nonlinear effects lead to the strongest spectral broadening. The asymmetry in the broadening for positive and negative linear chirps can be explained by a red-shifted rising front and a blue-shifted trail for positive chirps which leads to increased broadening in a nonlinear medium and vice versa less broadening for negative chirps.
It should be mentioned that, moreover, a small blue shift due to the onset of free electron plasma generation occurs for all pulse shapes \cite{wood91}. This shift depends on the laser pulse induced ionization rate for generating free electrons.        
\begin{figure}[t]
	\centering
  \includegraphics[width=0.8\textwidth]{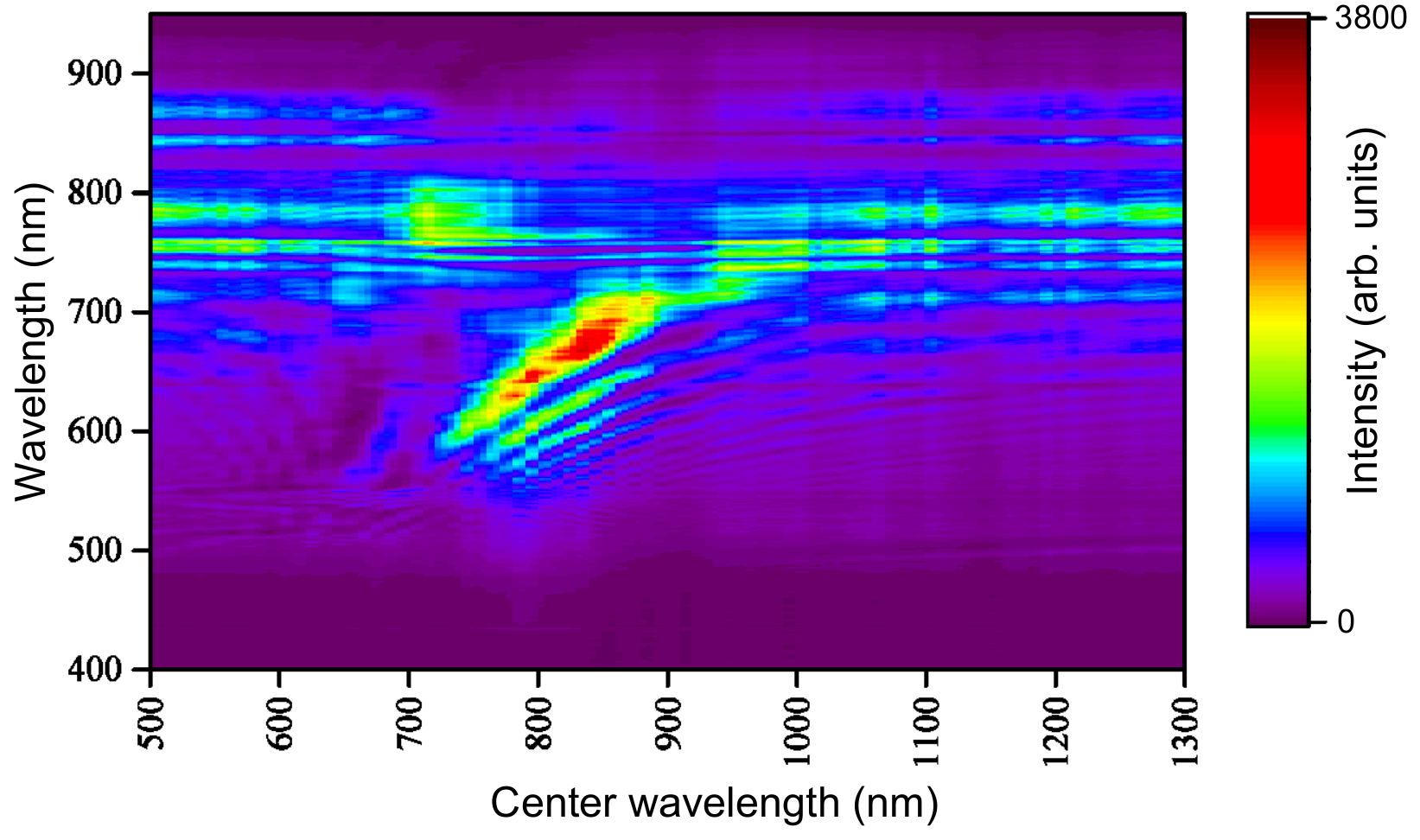}
	\caption{Experimental results for wavelength-shifted third order phase functions with a scaling factor of $b_3 = 2\cdot10^2 fs^3$. The x-axis denotes the center wavelength of the chirp expansion. With a decreasing antisymmetry wavelength a spectral band shifts accordingly to lower wavelengths.}
	\label{l0shift}
\end{figure}
\begin{figure}[ht]
	\centering
  \includegraphics[width=0.6\textwidth]{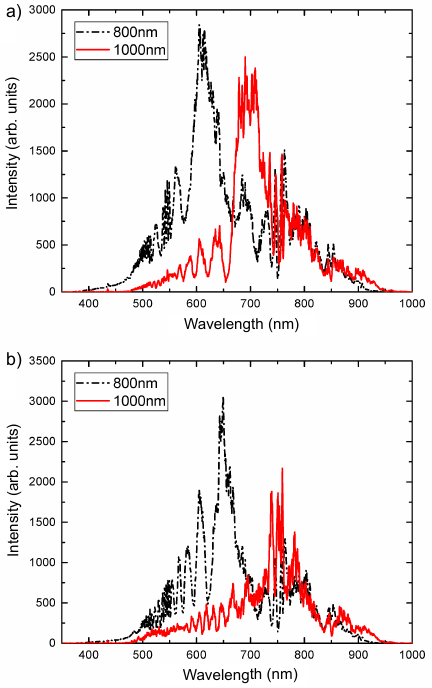}
	\caption{Recorded spectra for selected third order phase values and center wavelengths of expansion (see Fig. \ref{l0shift}). a) Spectra for pulses of $100\,fs^3$ third order phase factor and $\lambda_c$ at $800\,nm$ and $1000\,nm$, respectively.  b) Corresponding spectra for laser pulses of $200\, fs^3$ third order phase factor. In these cases, narrower spectral features and shifts to longer wavelengths are observed.}
	\label{l0shiftcut}
\end{figure}

As a next step 
to understand the  influence of pulse modulation we investigated the spectral dependence on the quadratic chirp (see Fig. \ref{qscan}). We observe  two dominant asymmetric spectral bands which shift towards higher frequencies for lower absolute chirp values. The differences in the spectra for positively and negatively chirped pulses originate from the temporal pulse asymmetry. For laser pulses with $b_3>0\,fs^3$ the pulse has a slow leading and a steep trailing flank while for $b_3<0\,fs^3$ the argon atoms experience a steep rise in the electric field strength followed by a long decrease. The slope determines the wavelength shift by self-phase modulation which is visible in Fig. \ref{qscan}. For positively quadratically chirped pulses, one observes a narrow band shifted far towards higher frequencies (due to the steep trailing flank) and a broad spectral band shifted to lower frequencies (resulting from the slow leading flank). These effects decrease for larger chirps because it is primarily dependent on the peak intensity.

Besides the shifted band we find a broad dip in the spectrum for positive quadratic chirps which broadens for pulses with lower quadratic chirp and higher peak intensities. This feature can be used to optimize intensity ratios between spectral regions of filament spectra when those are experimentally required. In order to see whether these features could be controlled in an easy parametric way, we employed a parametrization consisting of a quadratic chirp expanded around a selectable frequency $\omega_c$.
\begin{equation}
\varphi(\omega) = \frac{b_3}{6} (\omega-\omega_c)^3
\end{equation}
This is an antisymmetric phase function which was shown to optimize second and third order processes around the wavelength of antisymmetry \cite{pastirk_selective_2003}. The parametrization was chosen because the dominating $\chi^{(3)}$ processes in white light generation are expected to be controllable in a similar way.

The obtained spectral modifications in Fig. \ref{l0shift} can be understood by taking a closer look at the turning points of the 
third order phase function. At these points of antisymmetry ($\omega_c$) the first derivative of the phase (the instantaneous frequency) $\partial\varphi(t)/\partial t$ becomes 0. This implies that, at the peak intensity, frequency-components around $\omega_c$ are dominant. 
Hence, the spectral maxima in the $\lambda_c$
shift scans can be seen as beeing partially moved in correspondance to these antisymmetry points. The given explanation is supported by similar 
slopes of the center wavelength and the wavelengths of the maxima. It can further be observed that a shift of $\lambda_c$ to lower wavelengths results in a shift of the main peak to lower wavelengths.
By comparing  Fig. \ref{l0shiftcut} a) and b) it is obvious that larger quadratic chirps lead to sharper spectral features which can be explained by a spectrally narrower central zero phase region. These features can also be regarded as a demonstration of the feasible control of the nonlinear filament properties by multiphoton intrapulse interference effects \cite{pastirk_selective_2003} utilizing third order phase functions.  

A smaller shift of the spectral peak compared to the central wavelength is noticed for pulses with $b_3 = 200fs^3$ (Fig. \ref{l0shiftcut} b)) compared to the lower third order phase factor of $100 fs^3$ in Fig. \ref{l0shiftcut}. This observation of a minor maxima shift compared to the central wavelength is due to more dominating third order phases corresponding with reduced peak intensity.
In Fig. \ref{l0shiftcut} b) it is e.g. visible that for $\lambda_c = 1000\,nm$ the received less intense pulse generates only minor spectral modulations by self-phase modulation or ionization.
As a useful method for controlled spectral modification one can choose the desired spectral 
position by adjusting the center wavelength and one can select the spectral width by modifying the prefactor. 
This fast parametric spectral modulation can favorably be utilized for further applications. 
\begingroup
\squeezetable
\begin{table}[b]
\caption{Results of the optimizations for the given spectral ranges. A large difference is obtained between both solutions.} 
\vspace{0.4cm}
\begin{tabular}{c | c | c | c | c | c}
$(600-650)\,nm$ & $(650-700)\,nm$ & $\lambda_c$ & $b_2$ & $b_3$ & $b_4$\\
\hline
min & max & $631.9\,nm$ & $69.6\,fs^2$ & $1152\,fs^3$ & $4455\,fs^4$\\
max & min & $733.0\,nm$ & $-2.1\,fs^2$ & $386\,fs^3$ & $1422\,fs^4$\\
\end{tabular}
\end{table}
\endgroup

\begin{figure}[t]
	\centering
  \includegraphics[width=0.8\textwidth]{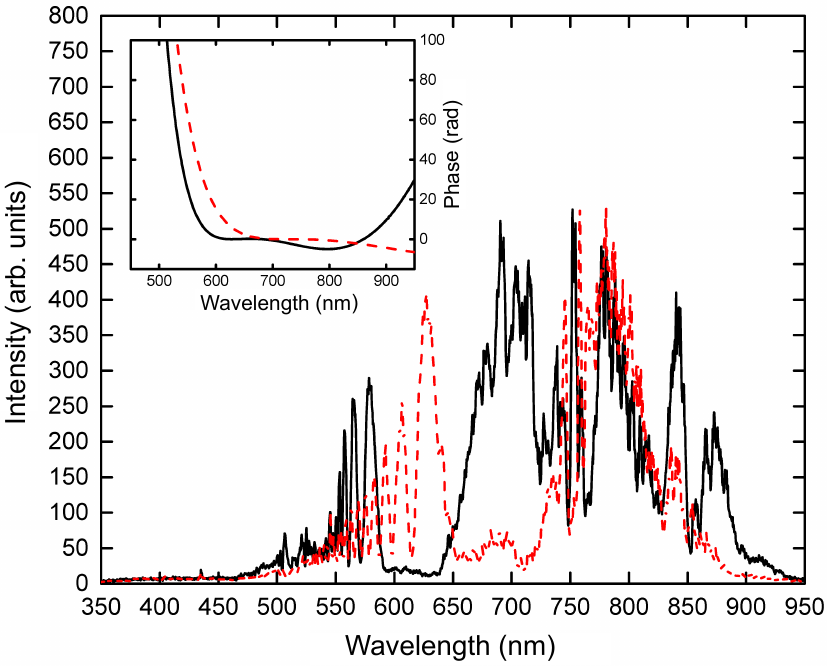}
	\caption{Optimizations of the ratios of spectral areas, where the area between $(600-650)\,nm$ was maximized and between $(650-700)\,nm$ was minimized (dashed red curve), and in contrast where $(600-650)\,nm$ was minimized and $(650-700)\, nm$ was maximized (solid black curve).}
	\label{specAreas}
\end{figure}

After understanding the basic effects of parametric pulse shaping on the generated filament spectra we employed an evolutionary algorithm \cite{Patas} to optimize a selected ratio of two spectral bands. The utilized parameter set consisted of a chirp expansion up to the third order around a center wavelength $\lambda_c$.
Table I shows the results of two optimizations either maximizing or minimizing the ratio of the integrated spectra between $600\,nm-650\,nm$ and $650\,nm-700\,nm$ and Fig. \ref{specAreas}  displays the corresponding spectra. If the spectral band of higher wavelengths is maximized the optimization proceeds for 27 generations of the evolutionary algorithm before converging to an intensity ratio of approximately 11.7. 
The most favorable optimization for the inverted fitness function yielded an intensity ratio of 4.1 after 33 generations, which corresponds to an overall modulation of the ratio by a factor of about 48. Each generation has 30 individuals including one survivor from the previous generation. This result reveals how evolutionary algorithms can be used in connection with parametric pulse shaping to find optimized pulse shapes for filamentation for various experimental applications.

The inset of Fig. \ref{specAreas} presents the optimized phase functions (without the offset phase used for compensation) which lead to the displayed spectra. The spectral features can be explained with the wavelengths of the inflection points which can be regarded as local antisymmetry points. The solid phase function for maximizing $650 - 700\,nm$ shows two local antisymmetry points whereas the dashed phase function shows one inflection point in between which corresponds to spectral maxima, respectively, as can be observed in Fig. \ref{specAreas}. Hence, the parametric optimizations utilize the spectral positions of local antisymmetry points to specifically modify the spectra.
This proves that higher order terms are relevant for precise spectral control.
We want to conclude that filaments can be well modified in a controlled way. Pulse shapes have an immense impact on the spectrum obtained from a filament and should be examined more closely in this regard in future. Laser pulse shaping with filamentation could become a versatile tool when looking for custom spectral shapes for further experiments.

\section*{Acknowledgements}
The authors thank Prof. Dr. L. W\"oste for his encouragement. A. L. acknowledges the Klaus Tschira Foundation (KTS) for financial support (project 00.314.2017).  M.M. acknowledges funding from MHV
fellowship grant number: PMPDP2-145444 and NCCR MUST Womens Postdoc Award.
S. H. acknowledges cofunding under FP7-Marie Skłodowska-Curie – NCCR MUST
program (200021-117810). The work was supported by the ERC advanced grant
"Filatmo" and the SNF NCCR MUST grant.

%


\end{document}